\documentclass[twocolumn,showpacs,preprintnumbers,amsmath,amssymb,superscriptaddress]{revtex4}
\usepackage{graphicx}% Include figure files
\usepackage{dcolumn}% Align table columns on decimal point
\usepackage{bm}% bold math
\def \ecm{$e\,$cm}
\def \figabbr{Figure\ }
\def \eqnabbr{Eq.\ }

\def \upperlimit{$\left| d_{n}\right| <2.9 \times 10^{-26}$ \ecm}
\def \gradient{$(1.57\pm0.08)\times10^{-26}$ \ecm/ppm}
\def \fitgrad{$(1.90\pm0.25)\times10^{-26}$ \ecm/ppm}

\def \vecB{$\mathbf{B}_0$}
\def \dnf{d_{n,Hg,f}}
\def \dBzdz{\partial B_z/\partial z}
\def \dnHgf{d_{n,Hg,f}}

\begin{document}

\title{An Improved Experimental Limit on the Electric-Dipole Moment of the Neutron
}
\author{C.A. Baker}\affiliation{Rutherford Appleton Laboratory, Chilton, Didcot, Oxon OX11 0QX, UK}
\author{D.D. Doyle}\affiliation{Department of Physics and Astronomy, University of Sussex, Falmer, Brighton BN1 9QH, UK}
\author{P. Geltenbort}\affiliation{Institut Laue-Langevin, BP 156, F-38042 Grenoble Cedex 9, France}
\author{K. Green}\affiliation{Rutherford Appleton Laboratory, Chilton, Didcot, Oxon OX11 0QX, UK}
\affiliation{Department of Physics and Astronomy, University of Sussex, Falmer, Brighton BN1 9QH, UK}
\author{M.G.D. \surname{van der Grinten}}\affiliation{Rutherford Appleton Laboratory, Chilton, Didcot, Oxon OX11 0QX, UK}\affiliation{Department of Physics and Astronomy, University of Sussex, Falmer, Brighton BN1 9QH, UK}
\author{P.G. Harris}\affiliation{Department of Physics and Astronomy, University of Sussex, Falmer, Brighton BN1 9QH, UK}
\author{P. Iaydjiev\footnote{On leave from Institute of Nuclear Research and Nuclear Energy, Sofia, Bulgaria}}\affiliation{Rutherford Appleton Laboratory, Chilton, Didcot, Oxon OX11 0QX, UK}
\author{S.N. Ivanov\footnote{On leave from Petersburg Nuclear Physics Institute, Russia}}\affiliation{Rutherford Appleton Laboratory, Chilton, Didcot, Oxon OX11 0QX, UK}
%\author{A.I. Kilvington}\affiliation{Rutherford Appleton Laboratory, Chilton, Didcot, Oxon OX11 0QX, UK}
\author{D.J.R. May}\affiliation{Department of Physics and Astronomy, University of Sussex, Falmer, Brighton BN1 9QH, UK}
\author{J.M. Pendlebury}\affiliation{Department of Physics and Astronomy, University of Sussex, Falmer, Brighton BN1 9QH, UK}
\author{J.D. Richardson}\affiliation{Department of Physics and Astronomy, University of Sussex, Falmer, Brighton BN1 9QH, UK}
\author{D. Shiers}\affiliation{Department of Physics and Astronomy, University of Sussex, Falmer, Brighton BN1 9QH, UK}
\author{K.F. Smith}\affiliation{Department of Physics and Astronomy, University of Sussex, Falmer, Brighton BN1 9QH, UK}

\date{\today}

\begin{abstract}
An experimental search for an electric-dipole moment (EDM) of  the neutron has been carried out at the Institut Laue-Langevin (ILL), Grenoble.  Spurious signals from magnetic-field fluctuations were reduced to insignificance by the use of a cohabiting atomic-mercury magnetometer.  Systematic uncertainties, including geometric-phase-induced false EDMs, have been carefully studied.   Two independent approaches to the analysis have been adopted. The overall results may be interpreted as an upper limit on the
absolute value of the neutron EDM of \upperlimit\ (90\% CL).
\end{abstract}

\maketitle

\section{Introduction}

Measurements of particle electric-dipole moments (EDMs) \cite{harris99,romalis01,regan02} are of significant interest because they provide some of the tightest constraints on extensions to the Standard Model, such as supersymmetry, that attempt to explain the mechanisms underlying CP violation \cite{peccei77,he88,barr93a,ramsey95,ellis96,lebedev04,pospelov05,abel06}.

%In order for particles to have electric dipole moments, the forces concerned in their structure must violate both parity (P) and time reversal (T) symmetries.  Although P violation is maximal in the weak interaction, T (and hence CP) violation is rare and its origins are not yet understood.  Extensions to the Standard Model, such as additional Higgs fields, right-handed currents or supersymmetric partners typically give rise to neutron EDM (nEDM) contributions which are of order 10$^{-25}$ to 10$^{-27}$ \ecm\ \cite{he88,ellis96,pospelov05,abel06}; dipole moments of this size might also come from CP violation in QCD \cite{peccei77}. Experimental measurements of particle EDMs \cite{romalis01,regan02,lebedev04}, and in particular that of the neutron, are providing some of the strongest additional constraints on these theories \cite{barr93a,ramsey95}.

This neutron-EDM experiment, and the performance of its cohabiting mercury magnetometer, have been discussed in earlier publications \cite{harris99,green98}.  The final result presented in this Letter incorporates a comprehensive analysis of systematic errors, some of which were undiscovered at the time of the earlier measurements.  

%The experimental method is summarised in Section II.  Section III discusses the geometric-phase (GP) effects that are the leading source of systematic error.  Section IV presents the data analysis, which involves characterisation of and compensation for these GP effects, and results.  Finally, Section V discusses other, less significant, potential sources of error.

\section{EDM measurement technique}

The measurement was made with ultracold neutrons (UCNs) stored in a trap permeated by uniform $%
\mathbf{E}$- and $\mathbf{B}$-fields. This adds terms $-\mathbf{\mu }_{n}\cdot 
\mathbf{B}$ and $-\mathbf{d}_{n}\cdot \mathbf{E}$ to the
Hamiltonian determining the states of the neutron. Given parallel $\mathbf{E}
$ and $\mathbf{B}$ fields, the Larmor frequency $\nu _{\uparrow \uparrow }$
with which the neutron spin polarization precesses about the field direction
is 
\begin{equation}
\label{eqn:hnu}
%h\nu _{\uparrow \uparrow }=2\mathbf{\mu }_{n}\cdot \mathbf{B}+2\mathbf{d}_{n}\cdot \mathbf{E.}
h\nu _{\uparrow \uparrow }=|2\mu_{n}B+2d_{n}E|.
\end{equation}
For antiparallel fields, 
%$h\nu _{\uparrow \downarrow }=2\mathbf{\mu }_{n}\cdot \mathbf{B}-2\mathbf{d}_{n}\cdot \mathbf{E}$.
$h\nu _{\uparrow \downarrow }=|2\mu_{n}B-2d_{n}E|$.
Thus the experiment aimed to measure any shift
in the transition frequency $\nu $ as an applied $\mathbf{E}$ field
alternated between being parallel and then antiparallel to $\mathbf{B}$.

\begin{figure}
\begin{center}
    \resizebox*{0.5\textwidth}{!}{\includegraphics[clip=true, viewport = 70 265 484 754]
    {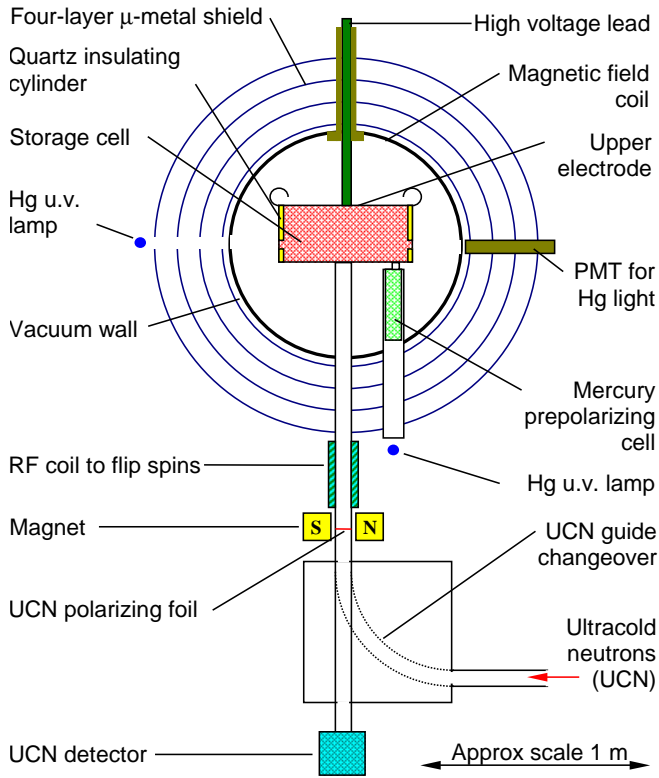}}
    \caption{(Color online) Experimental apparatus}
    \label{fig:edm_app}
\end{center}
\end{figure}

A schematic of the apparatus is shown in \figabbr\ref{fig:edm_app}. The UCNs were prepared
in a spin-polarized state 
%(polarization product $\sim0.6$) 
by transmission through a thin, magnetized iron
foil, and entered a cylindrical 21-liter trap within a 1 $\mu$T uniform vertical magnetic field \vecB.  

Approximately 20~s were needed to fill the trap with
neutrons, after which the entrance door was closed pneumatically. The
electric field, of approximately 10 kV/cm, was generated by applying high voltage to
the electrode that constituted the roof of the trap, while keeping the floor electrode grounded.
The electrodes were made of diamond-like-carbon coated Al, and the side wall was SiO$_2$.

The transition frequency $\nu $ of the neutrons was measured using the Ramsey
separated oscillatory field magnetic resonance method. During the storage
period, the neutrons interacted coherently with two 2 s
intervals of oscillating magnetic field having a chosen frequency close to
the Larmor frequency. The two intervals were separated by a period $T = 130$ 
s of free precession. The last step was to count the number
of neutrons $N_{\uparrow }$ and $N_{\downarrow }$ that finished in each of
the two polarization states. This was achieved by opening the entrance door
to the trap and allowing the neutrons to fall down onto the
polarizing foil, which then acted as a spin analyzer. Only those in the
initial spin state could pass through to the detector, which was a proportional
counter in which neutrons were detected via the reaction $n+^{3}$He $%
\rightarrow ^{3}$H$+p.$ During one half of the counting period, an r.f.\ magnetic
field was applied in the region above the polarizing foil; this flipped the
spins of the neutrons, thereby also allowing those in the opposite spin
state to be counted.  Each batch cycle yielded about 14,000 UCN counts.   Within a run the data-taking operations were cycled 
continuously for 1-2 days.  Periodically, after a preset number (normally 16) of batches,
the direction of $\mathbf{E}$ was reversed.  All other settings were held constant during a run. Every 10-20 runs, \vecB\ was reversed so that half of the full data set was taken with \vecB\ upwards and half with \vecB\ downwards. We adopt a system as in \cite{pendlebury04} where the $\hat{k}$ vector of our $z$ axis follows the direction of \vecB. Hence, $B_0$ is always positive, while the gravitational displacement of the UCNs changes sign. 

The magnetometer used the precession frequency of $I=1/2$ atoms of $^{199}$Hg ($3\times
10^{10}$ atoms/cm$^{3}$; $\mu_n/\mu_{Hg} = \gamma_n/\gamma_{Hg} = -3.842$) stored simultaneously in the same trap as
the neutrons.  Using \eqnabbr (\ref{eqn:hnu}) for both UCNs and Hg, and assuming that both experience the same $B$, we find that to first order in the EDMs $d$, 
\begin{equation}
\label{eqn:freq_ratio}
\frac{{\nu_n }}{{\nu _{Hg} }} = \left| {\frac{{\gamma _n }}{{\gamma _{Hg} }}} \right| + \frac{{(d_n  + \left| {\gamma _n/\gamma _{Hg} } \right|d_{Hg} )}}{{\nu _{Hg} }}E = \left| {\frac{{\gamma _n }}{{\gamma _{Hg} }}} \right| + \frac{{d_{meas} }}{{\nu _{Hg} }}E.
\end{equation}
For each run, $d_{meas}$ was obtained from a linear fit to the ratio $\nu_n / \nu_{Hg}$ versus $E$.  \eqnabbr (\ref{eqn:freq_ratio}) shows that $d_{meas}$ contains a contribution from $d_{Hg}$. The true $d_{Hg}$ has been shown to be $(-1.06\pm0.49\pm0.40)\times 10^{-28}$ \ecm \cite{romalis01}, so it introduces a systematic error of $(-0.4\pm0.3)\times10^{-27}$ \ecm\ into $d_{meas}$.% (see Table 1).

\section{Geometric-phase effects}

     To the true $d_n$ and $d_{Hg}$ within $d_{meas}$ there will also be added coefficients of fractional shifts in $\nu_n$ and $\nu_{Hg}$, from other causes, which are linear in $E$ and thus constitute additional systematic errors.  The most important of these involves a geometric phase (GP) arising when the trapped particles experience a gradient $\partial B_{0_z}/\partial z$ in the presence of $E$ \cite{pendlebury04}. Fortunately, the centre of gravity of our UCNs is $\Delta h = 0.28$ cm lower than that of the (warmer) Hg atoms, so an observed shift of $\nu_n / \nu_{Hg}$ away from $|\gamma_n/\gamma_{Hg}|$ gives a measure of the volume-averaged $\langle\partial B_{0_z}/\partial z\rangle_V$,  via the result (\eqnabbr 86 in \cite{pendlebury04})
\begin{equation}
 R_a = \left| \frac{\nu_n}{\nu_{\rm Hg}}\cdot\frac{\gamma_{\rm Hg}}{\gamma_n}\right| 
 = 1 \pm \Delta h \frac{\langle\partial B_{0_z}/\partial z\rangle_V}{B_{0_z}},
 \end{equation}
where the + sign corresponds to \vecB\ downwards.

     In this experiment the contribution of the GP effect in the Hg to $d_{meas}$ is 50 times larger than the GP effect of the UCNs.  Writing the GP false contribution to $d_{meas}$ from the Hg as $d_{n,Hg,f}$,  it is shown in \cite{pendlebury04} (\eqnabbr 87) that
\begin{equation}
\dnf = \pm\frac{\hbar}{8}|\gamma_n\gamma_{\textit{Hg}}|\frac{r_B^2B_{0_z}}{\Delta h\,c^2}\cdot(R_a-1) = \pm k\cdot(R_a-1),
\label{eqn:daf_Hgn}
\end{equation}
where $r_B$ is the trap radius and the + sign again corresponds to \vecB\ downwards.  It follows that we can write 
\begin{equation}
\label{eqn:d_meas}
d_{meas} = d^\prime_n + d_{{n,Hg,f}} = d^\prime_n \pm k\cdot(R_a-R_{a0}),
\end{equation}
where $d^\prime_n$ is the true $d_n$ plus all other systematic effects discussed below, and $R_{a0}$ is the value of $R_a$ where $\dBzdz = 0$.  \eqnabbr \ref{eqn:d_meas} defines two straight lines, one with positive slope for \vecB\ down and one with a negative slope for \vecB\ up.  The crossing point $(R_{a0}, d^\prime_n)$ provides an estimator of $d^\prime_n$ free of $\dnf$.

\begin{figure}[ht]
  \begin{center}
    \resizebox*{0.5\textwidth}{!}{\includegraphics
    [clip=true, viewport = 50 205 540 765]
    {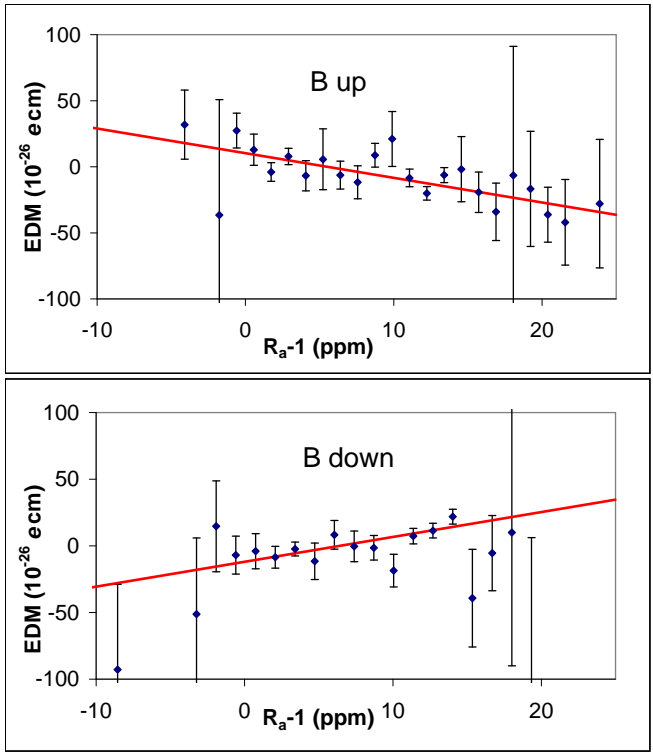}}
    \caption{(Color online) Measured EDM (binned data) as a function of the relative frequency shift of neutrons and Hg.}
    \label{fig:edm_vs_R}
  \end{center}
\end{figure}

Each run was made at a chosen value of $R_a$ by pre-adjusting currents in field-trimming coils.   \figabbr\ref{fig:edm_vs_R} shows the data (binned for clarity) for $d_{meas}$ as a function of $R_a$ for each direction of \vecB.  The lines represent a least-squares fit to all 554 of the (unbinned) run results, using as free parameters the two intercepts and a common absolute slope $k$.  This yields $\chi^2/\nu$ = 652/551 and $k =$ \fitgrad, which is within 1.3$\sigma$ of the expected value of \gradient\ from \eqnabbr (\ref{eqn:daf_Hgn}).  
The slope $k$ can be altered by a few percent (although still remaining highly symmetric under \vecB\ reversal) by various mechanisms including the UCNs' own GP signal (a 2\% effect); uncertainty in $\Delta h$ (4\%); a slight reduction in mean free path due to cavities and grooves in the electrodes as well as to the presence of 10$^{-3}$ torr of He gas to prevent sparks \cite{pendlebury04, lamoreaux05} (1\%); and asymmetric surface relaxation of the Hg (up to 5\%).

\subsection{Anomalous GP effects}

There are some processes that can interfere with the above GP error removal -- essentially any process that changes $R_a$ and/or $\dnf$ without conforming to the ratio between the two given by \eqnabbr (\ref{eqn:daf_Hgn}), and where, in addition, the changes differ with the direction of \vecB. 

    First, there are several processes that shift $R_a$ but not $\dnf$. Changing $\gamma_n / \gamma_{Hg}$, for example, shifts the two lines of \figabbr 2 in the same direction by the same amount, leaving $d^\prime_n$ unaffected but changing $R_{a0}$. We note at this point that our final $R_{a0}$ is consistent with the $\gamma_n / \gamma_{Hg}$ value from the literature (1$\sigma \equiv 1.7$ ppm) \cite{greene79,cagnac61}, after allowing for our observed $B_x$ and $B_y$ fields with finite $\partial B_x/\partial y$ and/or $\partial B_y/\partial x$ but with ($\partial B_x/\partial x + \partial B_y/\partial y) = 0= -\dBzdz$ (e.g., a quadrupole aligned with z, with $B_x = qy, B_y = qx$), which cause $R_a$ to increase \cite{pendlebury04} without contributing to $\dnf$. Any change in such fields when \vecB\ is reversed can result in a differential shift of all $R_a$ values, and thus of the two lines, thereby changing $R_{a0}$ and $d^\prime_n$. Below, we describe our measurements of the differential shift in $R_{a0}$ and state the correction to $d^\prime_n$. 

    $B$-field averaging in the trap is affected by localised loss of UCN and Hg particles, and by polarization loss in the presence of the $10^{-3}$ fractional $B_0$ inhomogeneities, which may change with \vecB\ direction. However, we estimate that the resulting $R_a$ shifts are $< 0.1$ ppm and $< 0.01$ ppm for the UCN and Hg respectively, and that they will be indistinguishable from the quadrupole shifts. 

     Light shifts \cite{cohen-tannoudji72, corney} in $\nu_{Hg}$ will shift $R_a$. They are produced by any small component, parallel to \vecB, of the $^{204}$Hg probe light beam passing through the precessing $^{199}$Hg atoms. This component, and the $R_a$ shift, reverse sign on reversal of \vecB. A slight dependence of $R_a$ on the incident light intensity was indeed found, the magnitude $\sim 0.2$ ppm being in agreement with theory. A correction to $d_{meas}$ was made on a run-by-run basis, leading to an overall correction of $(3.5 \pm 0.8) \times 10^{-27}$ \ecm.

   Second, there are processes that generate an enhanced $\dnHgf$. 
   The field of a permanent magnetic dipole (PMD) close to the trap makes a non-uniform $\dBzdz$ and adds an (enhanced) GP $d_{dip}$ \cite{harris06} to $\dnHgf$, but shifts $R_a$ in accord with \eqnabbr (\ref{eqn:freq_ratio}) and in opposite senses for the two \vecB\ directions. The two changes are in a ratio greater than that given by \eqnabbr (\ref{eqn:daf_Hgn}). This shifts the lines of \figabbr 2 upwards, adding $(d_{dip}-d_4)$ to $d^\prime_n$, where $d_4$ is the prediction of \eqnabbr (\ref{eqn:daf_Hgn}) for the $\langle\dBzdz\rangle_V$ of this PMD. Our fluxgate magnetometer surveys of the trap cannot rule out PMD fields of less than 1 nT at 2 cm from the inner surface. Large areas of the trap are SiO$_2$ or Al, backed by large voids, and do not come under suspicion; but the Hg and UCN doors involve a heterogeneous collection of small parts close to the trap. We allow a $d^\prime_n$ uncertainty of $\pm 6.0 \times10^{-27}$ \ecm\ to allow for an undetected 1 nT PMD at the Hg door.  In the case of the UCN door we have better diagnostics. The trap used when taking EDM data has a small cavity in the lower electrode, 4.0 cm deep and 6.8 cm in diameter, sealed from below by the door. A PMD in the door mechanism can contribute a $\dBzdz$ field to the cavity and to the rest of the trap. The field in the cavity contributes a strong shift in $R_a$, while contributing negligibly via the GP to $\dnHgf$  due to the small cavity radius. As a result, the lines of \figabbr 2 are shifted in opposite directions, again adding a systematic error to $d^\prime_n$.

\subsection{Auxiliary measurements}

      Our additional diagnostics came from separate $\nu_n$, $\nu_{Hg}$  and $R_a$ measurements (without $E$ fields) in two auxiliary traps having roofs that could be raised or lowered to change the height $H$. The traps were built on the same lower electrode and door mechanism that were used for EDM data taking. Assuming $B_z(z) = b_0 + b_1z + b_2z^2$, one can show that $\langle\dBzdz\rangle_V$ = 0 for a trap height $H$ when $\nu_{Hg}$ is the same for roof settings at $H/2$ and $H$. With such a field established, $R_{a0}$ was measured for several heights $H$. The resulting forms of $R_{a0}(H)$ for \vecB\ up and \vecB\ down led to the conclusion that there was a dipole field of strength $\sim 1$ nT penetrating into the door cavity. They also confirmed the presence of quadrupole fields, which the fluxgate scans showed vary little with $H$. Polarization data from the EDM runs further substantiated this:  UCNs of different energies have different $\Delta h$s, so the surviving UCN polarization decreases in proportion to $(\partial B_z\partial z)^2$, and the values of $R_{a0}$ where the ${\bf B}_0$ up and down polarizations peak are in excellent agreement with the auxiliary trap results.
   
    The first auxiliary trap used had a smaller radius, 18.5 cm rather than 23.5 cm, and a cavity depth of 6.0 rather than 4.0 cm. These differences amplify the $R_a$ shifts from a dipole field in the cavity by 1.50 and reduce the quadrupole shifts by a factor 1/1.8. Our systematic error correction to  $d^\prime_n$ to allow for the combined door dipole and quadrupole fields is $(0.69 \pm 0.28) \times 10^{-26}$ \ecm. The auxiliary trap used to measure $\Delta h$ by obtaining $R_a$ as a (linear) function of a series of known $\langle\dBzdz\rangle_V$ was made as similar to the data-taking trap as possible in dimensions and materials so as to reproduce the same UCN velocity spectrum.

\begin{table}
	\begin{center}
		\begin{tabular}{|l|l|l|}
		\hline
		{\bf Effect}  & {\bf Shift}& {\bf $\sigma$}\\
		\hline 
		Door cavity dipole                         & -5.6 & 2.0   \\
		\hline
		Other dipole fields                        & 0.0 & 6.0 \\
		\hline
		Quadrupole difference                   & -1.3 & 2.0 \\
		\hline
		${\bf v}\times{\bf E}$ translational 	     & 0.0 & 0.03\\
		\hline
		${\bf v}\times{\bf E}$ rotational          & 0.0 & 1.0\\
		\hline
		Second-order ${\bf v}\times{\bf E}$        & 0.0 & 0.02\\
		\hline
		$\nu_{Hg}$ light shift (geo phase)         & 3.5 & 0.8\\
		\hline
		$\nu_{Hg}$ light shift (direct)            & 0.0 & 0.2\\
		\hline
		Uncompensated B drift                      & 0.0 & 2.4\\
		\hline
		Hg atom EDM                                & -0.4 & 0.3\\
		\hline
		Electric forces                            & 0.0 & 0.4\\
		\hline
		Leakage currents                           & 0.0 & 0.1\\
		\hline
		AC fields                                  & 0.0 & 0.01\\
		\hline
		{\bf Total}                                & {\bf -3.8} & {\bf 7.2}\\
		\hline
		\end{tabular}
	\end{center}
\caption{Summary of systematic errors and their uncertainties, in units of 10$^{-27} ~e$cm. }
\end{table}

\section{Other systematic errors}

   We now consider the systematic errors not involving GP.
   
\subsection{${\bf v}\times{\bf E}$ effects} 

If the UCN have a net translational motion, any perpendicular component of the
${\bf E}$ field will be seen in their rest frame as a combination of ${\bf E}$
and ${\bf B}$ fields.  During the 130 s Ramsey measurement period, the UCN ensemble may warm slightly due to vibrations.  This causes the center of mass to rise by up to 0.1 cm.  If the volume-averaged angles between ${\bf E}$, ${\bf B}$ and ${\bf v}$ are each as high as 0.05 radians, the induced false EDM will be $0.03\times10^{-27}$ \ecm.

In a similar manner to the translational effect, any net rotational flow of the UCN in conjunction with a radial component of the ${\bf E}$ field may lead to an induced EDM signal.  However, any such flow of UCN is expected to be
attenuated by wall collisions before the first Ramsey pulse is applied. We
calculate that the maximum error to be expected from this source and from
higher-order ${\bf v}\times {\bf E}$ effects is below $1\times 10^{-27}$ \ecm.

\subsection{Direct light shift}

Analysis of the data suggested a possible small correlation between the intensity of the Hg reading light and the value of the applied ${\bf E}$ field.  Through the light shift, this could directly create an EDM signal.  However, to within its uncertainty, the dependence of $R_a$ on the light intensity has been removed.  The residual systematic uncertainty from this source is $0.2\times10^{-27}$ \ecm.

\subsection{Uncompensated magnetic field fluctuations}
There may also be residual effects from ${\bf B}$ field fluctuations.  For example, a dipole-like field ${\bf B}_d$ originating from the $\mu$-metal in the region of the HV feedthrough would be sensed by both neutrons and Hg, but with a difference given by $\delta B_d/B_d = 3\Delta h/r$ where $r\sim 55$ cm is the distance from the source of the field to the center of the bottle.  Thus, fluctuations in ${\bf B}$ that are correlated with the HV can be expected to be compensated up to a factor of about 70.  In order to study this, the Hg and neutron channels were analysed independently.  The neutrons yielded an EDM signal of $(17\pm4)\times10^{-26}$ \ecm; the Hg, once the GP contribution (as calculated from the average $R_a-1$ at which the data were taken) was subtracted, yielded $(-3.9\pm0.8)\times10^{-26}$ \ecm. These results are consistent with a common source of magnetic fluctuations correlated with the HV.  We therefore expect the Hg compensation to shield us from this systematic effect to a level of $17\times10^{-26}/70 = 2.4\times10^{-27}$ \ecm.

\subsection{Electric forces}
Another possible source of systematic error arises from electrostatic forces, which may move the electrodes slightly.  In conjunction with a magnetic field gradient, an HV-dependent shift in the ratio would then appear.  This was sought by looking for an EDM-like signal but with a frequency shift proportional to $|{\bf E}|$ instead of ${\bf E}$.  The $|{\bf E}|$ signal was consistent with zero, with an uncertainty of $4\times10^{-26}$ \ecm. If the HV magnitudes were slightly different for the two signs of ${\bf E}$, this effect would generate a false EDM signal.  Study of the measured HV and of the charging currents show that the HV magnitude was the same for both polarities to within about 1\%.  This systematic uncertainty is therefore $0.4\times10^{-27}$ \ecm.

\subsection{Leakage currents and sparks}

Analysis of the EDM as a function of leakage current shows no
measurable effect.  Leakage currents are typically of order 1 nA.  If this current were to travel 10 cm azimuthally around the bottle, the resultant ${\bf B}$ field would result in an apparent EDM of $0.1\times10^{-27}$ \ecm.  

\subsection{HV AC ripple}

AC fields are another possible cause of concern.  There is no differential ripple visible on the HV at the level of a few volts.  Sampling is done at 5 Hz with a bandwidth of 20 kHz, so any 50 Hz ripple would show up as beats.  This is certainly absent at the level of 50 V, which would give a false EDM of $0.01\times10^{-27}$ \ecm.

Low-frequency AC fields were sought by means of a pickup coil in conjunction with a phase-sensitive detector.  Shifts in $R_a$ from this source at the level of 0.02 ppm could not be ruled out. Cancellations in the corresponding EDM signal from reversals of the electric and magnetic fields would reduce any net contribution to below the level of $0.01\times10^{-27}$ \ecm.

\section{Analysis and results}

Two approaches were adopted in studying the data.  In the first, more straightforward, analysis, only the 293 runs with an uncertainty on $R_a$ of less than 0.05 and with -10 ppm $< (R_a-1)<$ 21 ppm  were accepted.  As the two resulting average $R_a-1$ values (8.948, 8.943 ppm for \vecB\ up, down respectively) were almost identical, with approximately equal amounts of data in each field direction, a  simple weighted average of the $d_{meas}$ data was used as an estimator of $d_n$.  The value obtained was
\[d_n = (-0.2 \pm 1.6 {\rm ~(stat)})\times10^{-26} ~e{\rm\,cm} ~(\chi^2/\nu = 1.24),\]
with an additional uniformly distributed systematic uncertainty of $\pm 1.2\times10^{-26}$ \ecm\ allocated to it in order to accommodate any potential systematic biases arising from the effects listed in Table 1.  The resulting distribution of possible values implies an upper limit of $|d_n| < 2.9\times10^{-26}$ \ecm\ (90\% CL).

The second analysis began with the \figabbr \ref{fig:edm_vs_R} fitted-lines crossing-point value $d^\prime_n = (-0.55\pm 1.51)\times10^{-26}$ \ecm\ (which includes the run-by-run light shift correction) and then applied the systematic-error corrections given in Table 1.  The final result from this approach is 
\[ d_n = \left(+0.2 \pm 1.5 ~{\rm (stat)} \pm 0.7 ~{\rm (syst)}\right)\times10^{-26} ~e{\rm\,cm}.\]
In this analysis, the systematic uncertainty is normally distributed.  The result implies an upper limit of $|d_n| < 2.8 \times10^{-26}$ \ecm\ (90\% CL).

\section{Conclusion}

The data set analysed here, which excludes data that have already been published \cite{harris99}, incorporates all neutron EDM measurements undertaken between the autumn of 1998 and the end of 2002.  The results overall may be interpreted as an upper limit on the absolute value of the neutron EDM of \upperlimit\ (90\% CL).

%\section{Acknowledgments}

\acknowledgments
The authors are grateful to N.F.\ Ramsey for many useful discussions; to Y.\ Chibane, M. Chouder, and I.A. Kilvington for their contributions during the development period; to E.N.\ Fortson's group for information and components relating to the atomic mercury magnetometry; and to R.\ Baskin for computer simulation work. The ILL has been generous in provision of neutron facilities.  The work was funded by grants from the UK Particle Physics and Astronomy Research Council..  Support from the RFFI, via 
Grant No.03-02-17305, is gratefully acknowledged by S.N.I.

\end{document}